\documentclass[11pt,titlepage]{article}
\usepackage{setspace} 
\usepackage{amsmath}
\usepackage{graphicx}
\usepackage[round,numbers,sort&compress]{natbib} 
\title{Multifrequency Forcing of a Hopf Oscillator Model of the Inner Ear}% Force line breaks with \\

\author{K.A.~Montgomery \thanks{
           Corresponding author.  Address: 
           Mathematics Department,
	   University of Utah,
	   155 South 1400 East, Room 233,
	   Salt Lake City, UT~84112, U.S.A.,
	   Tel.:~(801)419-1520 }
\\
	   Mathematics Department, \\
	   University of Utah, Salt Lake City, UT
	   }

\date{\today}

\begin{document}

\maketitle

\abstract{
In response to a sound stimulus, the inner ear emits sounds called otoacoustic
emissions. While the exact mechanism for the production of otoacoustic 
emissions is not known, active motion of individual hair cells is thought
to play a role. Two possible sources for otoacoustic emissions, 
both localized within individual hair cells,
include somatic motility and hair bundle motility.  Because physiological models 
of each of these systems are thought to be poised near a Hopf bifurcation, the dynamics
of each can be described by the normal form for a system near a Hopf bifurcation.   Here
we demonstrate that experimental results from three--frequency suppression experiments can be predicted 
based on the response of an array of noninteracting Hopf oscillators tuned at different
frequencies. This supports the idea that active motion of individual hair cells contributes to active processing
of sounds in the ear. Interestingly, the model suggests an explanation for differing results
recorded in mammals and nonmammals.  
}

\clearpage

\section*{Introduction}
\label{sec:intro}

The inner ear is more than a passive recorder of sounds.  It also actively
processes sounds using metabolic energy to spectrally analyze
and amplify the stimulus~\cite{Kandel,HM94,D92,FRH01,M01}.  One 
consequence of the inner ear's active sound processing 
is that it produces sounds called otoacoustic emissions.
Otoacoustic emissions, which consist of combinations of sounds at 
discrete frequencies, can occur either in the absence or in the presence of a
sound stimulus~\cite{PLM91}. The exact mechanism responsible for the active processing of sounds 
and the related production of otoacoustic emissions within 
the ear is not well known~\cite{HM94,FRH01,RR01}. Recording the emissions spectrum provoked by 
a stimulus provides a way to probe the physiological 
systems responsible for active processing of sound.

In nonmammals, active sound processing is thought to occur within individual hair cells~\cite{SN99,M03}.
Hair cells are mechanotransduction cells responsible for translating sound-induced mechanical motion
into an electrical signal that is received by the auditory nerve~\cite{Kandel,RR01}.  Each hair cell consists 
of a cell body which is contacted by the auditory nerve and a bundle of 
actin-supported fibers called stereocilia.  When sound stimulates the auditory organ,
the hair bundle is set into motion, causing transduction channels to be mechanically 
pulled open.  Potassium ions  flow through the transduction channels depolarizing
the cell and ultimately causing the firing of the auditory nerve.  In nonmammals,
each hair cell responds preferentially at a specific frequency, a quality that
makes the hair cell a prime suspect in the search for the source of the discrete-frequency otoacoustic
emissions.

Active motion of the hair bundle is considered to be a possible mechanism
for active sound processing in both the mammalian and nonmammalian 
ear~\cite{MG97,MKKY01,HCMM00,MMH00,MH99,RCF02,MBCH03,JMH93}.  
Experiments have shown that the hair bundle responds 
with more energy than the stimulus energy if stimulated near its resonance frequency~\cite{cmh98}.
It has been proposed that when the hair bundle is displaced, calcium enters through
the transduction channels and binds to a site inside the hair bundle~\cite{CC06,cmh98}. 
This binding causes a change in the
tension of the transduction channels which results in the motion of the hair bundle.  
In mammals, there is another source of active hair cell motion.  In 
response to depolarization, the cell bodies of outer hair cells contract due
to the action of the protein prestin~\cite{KBAF86,A87,ZSHLMD00}.  

Either the hair bundle motility or the outer hair cell somatic motility 
could be involved in the production of otoacoustic emissions.  Interestingly, 
a physiologically--based model for hair bundle motion 
has been shown to be poised near a Hopf bifurcation for physiologically 
reasonable parameters~\cite{MH99}.  The motion of the outer hair cells
also displays a resonance response~\cite{RACPDCB05} that is suspected 
to arise from a physiological system that is tuned near a Hopf bifurcation~\cite{ODI03}.

Assuming both the hair cell bundle motion and the outer hair cell motion 
is produced by a system poised near a Hopf bifurcation, the dynamics either
system can be described by the normal form for a system near a Hopf bifurcation~\cite{W1990},

\begin{equation}
 \frac{dA}{dt}=(a+ib)A-(c+id)|A|^2A. 
\label{eq:NF}
\end{equation}

\noindent The response properties of Eq. \ref{eq:NF} have been shown to reproduce qualitatively many of 
the amplification and tuning properties of the inner ear~\cite{EOCHM00,CDJP00}.  

The otoacoustic emissions produced by the ear in response to multifrequency stimuli
provide ample data concerning the active processing properties of the inner ear~\cite{PLM91}.
Here, we consider the predictions of the Hopf oscillator
model for three--frequency forcing experiments.  
It is of interest to determine whether observed otoacoustic 
emissions can be explained by an array of Hopf oscillators, each modeled by Eq. \ref{eq:NF} and,
if so, whether coupling between the motion of the oscillators is required 
to obtain observed otoacoustic emissions results.  We find that an array of noninteracting
Hopf oscillators, perhaps describing the motion of the hair bundles or outer hair cells, 
is adequate to qualitatively explain the results of the three--frequency forcing experiments 
in both nonmammals and mammals.

\section*{Analysis}
\label{sec:analysis}

Assuming both the motion of the hair 
bundle and the motion of the outer hair cell body can be modeled by a system tuned
near a Hopf bifurcation, the dynamics of each can be described by the normal form for
a system near a Hopf bifurcation, Eq. \ref{eq:NF}. In the normal form, 
the parameter $a$ is a measure of proximity to the bifurcation point.  When $a$
is small in magnitude and negative, the `cell' is tuned slightly below the Hopf bifurcation and responds 
to brief disturbances with decaying oscillations.  If $a$ is greater
than zero, the  `cell' is tuned above the Hopf bifurcation and the hair bundle
oscillates spontaneously.  The parameter $b$ is the natural frequency
of the cell at the onset of oscillation and $d$  is a measure of the shift in the frequency of the cell as the response
amplitude increases.  The parameter $c$ determines whether the system is supercritical ($c>0$) or
subcritical ($c<0$).  Here, we will concentrate on the supercritical case because
it allows for small-amplitude, spontaneous oscillations near the bifurcation point
similar to the spontaneous hair bundle oscillations that are observed experimentally~\cite{CF85}. 
If a small time-dependent forcing 
is applied the system~\cite{DMN02,CDJP00,MSS06}, the normal form must be modified to include a forcing term, $F$,  

\begin{equation}
 \frac{dA}{dt}=(a+ib)A-(c+id)|A|^2A +F.
\label{eq:NormalFormwF}
\end{equation}

\noindent In the case of single--frequency forcing, $F=fe^{i\omega t}$, the system can be 
analyzed by considering hair bundle motions responding
at the same frequency as the forcing frequency.
Substituting $A=R e^{i \omega t+ i \phi}$ into Eq. \ref{eq:NormalFormwF} yields the following
simple relationship between forcing amplitude and response amplitude,

\begin{equation}
  (a R - c R^3)^2+([b-\omega]R -d R^3)^2=f^2
\label{eq:algebraicNF}
\end{equation}

\noindent Using this relationship, Egu\'{\i}luz et al. and Camalet et al.~\cite{EOCHM00,CDJP00} 
each demonstrated that a generic system poised near a Hopf bifurcation displays 
many of the amplification and tuning properties
that are observed in the auditory system.  

Two-frequency forcing experiments have been useful in studying
the properties of otoacoustic emissions and determining their 
source.  In suppression experiments~\cite{RRR92,FG63},
the cochlea is stimulated 
by a primary tone as well as a second softer 
tone, referred to as a suppressor tone.  The addition of 
the softer tone has an effect on the magnitude
of the cochlear response at the primary frequency.  Specifically,
as the frequency of the suppressor tone approaches the frequency of
the primary tone, the magnitude of the component of the otoacoustic emission at 
the primary tone decreases.  The biological interpretation of this
is that since the maximum suppression occurs when the suppressor 
tone is near the primary frequency, it is likely that the otoacoustic
emission originates near the part of the cochlea tuned at the primary frequency.
Analysis of a Hopf oscillator tuned at the primary frequency
and forced by a primary and suppressor tone supports the biological interpretation.
Recently, Stoop et al.~\cite{SK04}, by analyzing a Hopf oscillator model 
showed that the effect of adding a second frequency close to
the primary frequency is to increase the effective damping
of the oscillator's response at the primary frequency.  Thus
a single cell, tuned near a Hopf bifurcation
point and near the primary frequency is adequate to reproduce the main
qualitative features of two-frequency suppression experiments.  

When the ear is stimulated by sound 
containing a linear combination of two primary frequencies $\omega_1$ and $\omega_2$,
the otoacoustic emissions spectrum is more complicated to analyze because distortion product 
otoacoustic emissions (DPOAE's) occur at linear combinations of the stimulus 
frequencies~\cite{PLM91,MK93,MD05}. In experiments, the largest DPOAE response
is observed to occur at the $2 \omega_1-\omega_2$ and $2 \omega_2 -\omega_1$ frequency components.
The presence of DPOAE's allows for more complicated 
multifrequency forcing experiments in which the amplitudes of the 
distortion products are considered.  For instance, suppression experiments can be performed in
which the cochlea is stimulated at a combination of two primary frequencies as well
as a smaller amplitude suppressor tone.  Then the effect of the suppressor
tone on the response at each of the primary frequencies and the distortion product 
frequencies can be recorded.   

In nonmammals, multifrequency forcing experiments, including two 
primary frequencies $\omega_1$ and $\omega_2$ ($\omega_1<\omega_2$) and a suppressor frequency, indicate that 
maximum suppression of the $2\omega_1-\omega_2$ distortion product frequency 
occurs when the suppressor tone is near the $\omega_1$ frequency~\cite{TM98,KM93,KMS95}.
Oddly, in mammals, the reverse trend is observed and maximum suppression of $2 \omega_1-\omega_2$ occurs
when the suppressor frequency is near the $\omega_2$ frequency~\cite{BK84,KJA98}.  If active
hair cell motion is responsible for the production of otoacoustic 
emissions, there must be an explanation for the discrepancy between emissions
in mammals and nonmammals.

Here, we consider the response properties of a Hopf oscillator under 
three-frequency forcing, $F=F_1 e^{i \omega_1 t}+F_2 e^{i \omega_2 t} + F_3 e^{i \omega_3 t}$.
Because the system is nonlinear, the response contains an infinite
number of frequencies, a small number of which will be represented prominently.
If one substitutes 
$A=A_1 e^{i \omega_1 t}+ A_2 e^{i \omega_2 t}+ A_3 e^{i \omega_3 t}$ 
into the nonlinear term from the normal form, $|A|^2A$, the result 
contains only certain frequency combinations.  
We will assume that those frequencies dominate the response, 
and thus consider a response, $A$, that is a linear combination
of those frequency components,

\begin{eqnarray} 
A &= &R_1 e^{i \omega_1 t+i \phi_1 }+R_2 e^{i\omega_2 t+i \phi_2}+R_3 e^{i \omega_3 t+i \phi_3} \nonumber \\
&&+R_{112} e^{i (2 \omega_1-\omega_2) t+i \phi_{112}} +R_{221} e^{i (2\omega_2-\omega_1) t+i \phi_{221}} \nonumber \\
&&+R_{113} e^{i (2 \omega_1-\omega_3 ) t+i \phi_{113}}+R_{223} e^{i (2\omega_2-\omega_3) t +i \phi_{223}} \nonumber \\
&&+R_{332} e^{i (2 \omega_3-\omega_2) t+i \phi_{332} } +R_{331} e^{i (2\omega_3-\omega_1)t+i \phi_{331}}  \nonumber \\
&&+R_{123} e^{i (\omega_1+\omega_2-\omega_3) t+i \phi_{123}}+R_{231} e^{i (\omega_2+\omega_3-\omega_1) t+ i \phi_{231}} \nonumber \\
&&+R_{312} e^{i (\omega_3+\omega_1-\omega_2) t+i \phi_{312}}  \label{eq:Asubs1} \end{eqnarray}

Substituting Eq. \ref{eq:Asubs1} into Eq. \ref{eq:NormalFormwF}
yields algebraic expressions relating the response amplitudes, response phases, and
the forcing amplitudes, $F_1$, $F_2$, and $F_3$. Frequency components not 
represented in Eq. \ref{eq:Asubs1} are neglected.  
This provides an analytical 
description for the response of the Hopf
oscillator to three-frequency forcing.

\begin{figure}
\centerline{\resizebox{3 in}{!}{\includegraphics{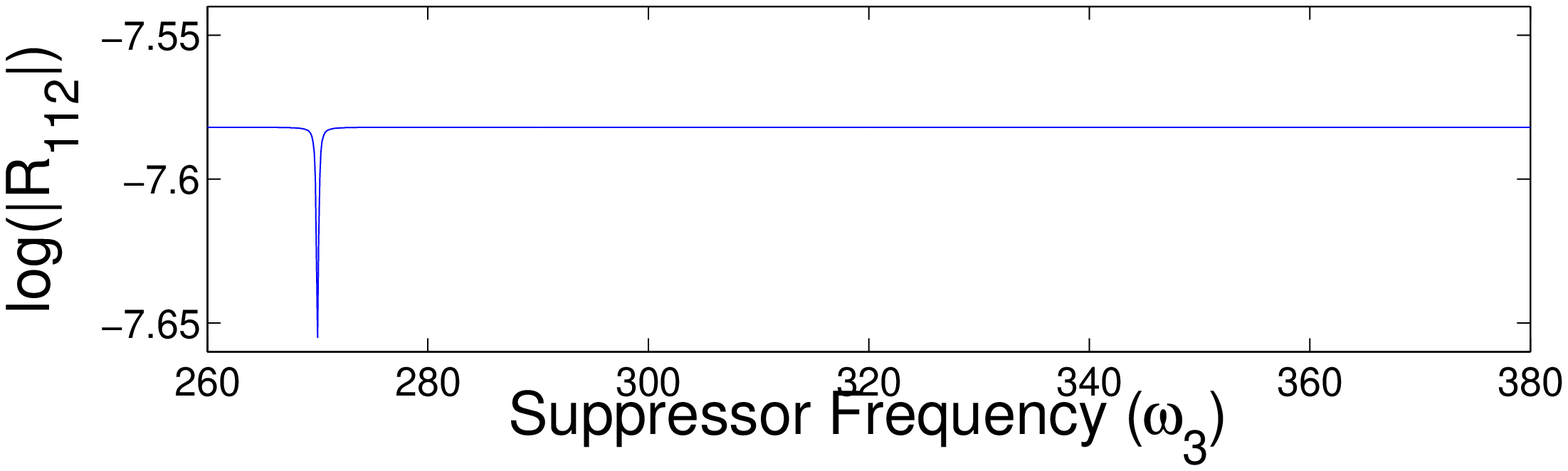}}}

\centerline{(a)}

\centerline{\resizebox{3 in}{!}{\includegraphics{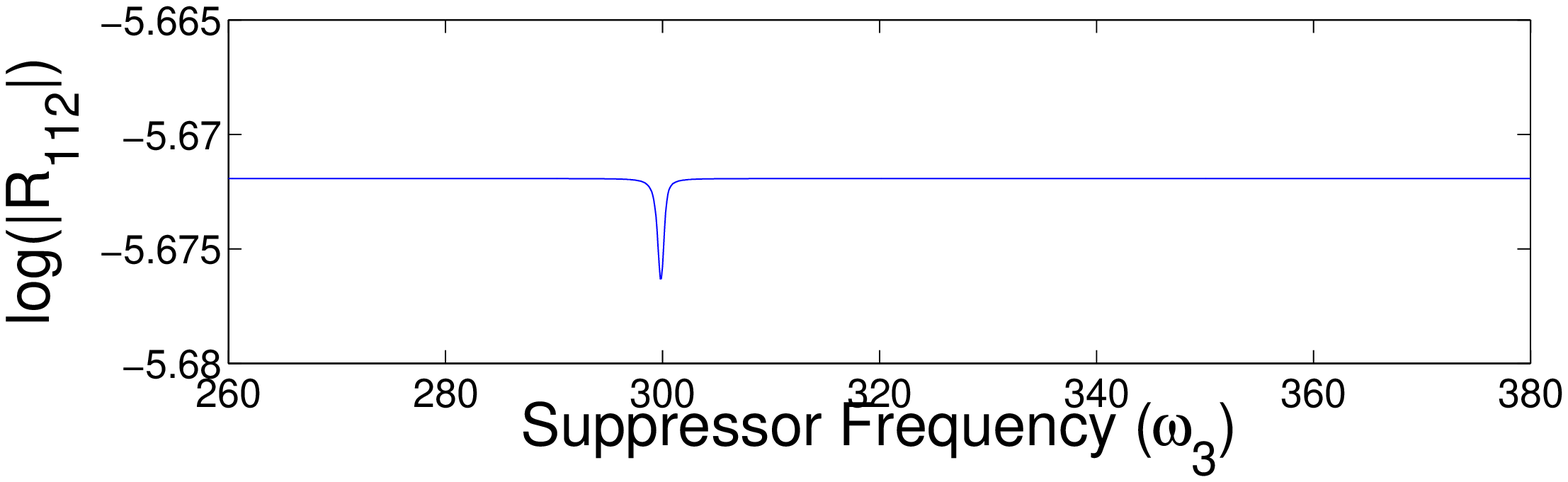}}}

\centerline{(b)}

\centerline{\resizebox{3 in}{!}{\includegraphics{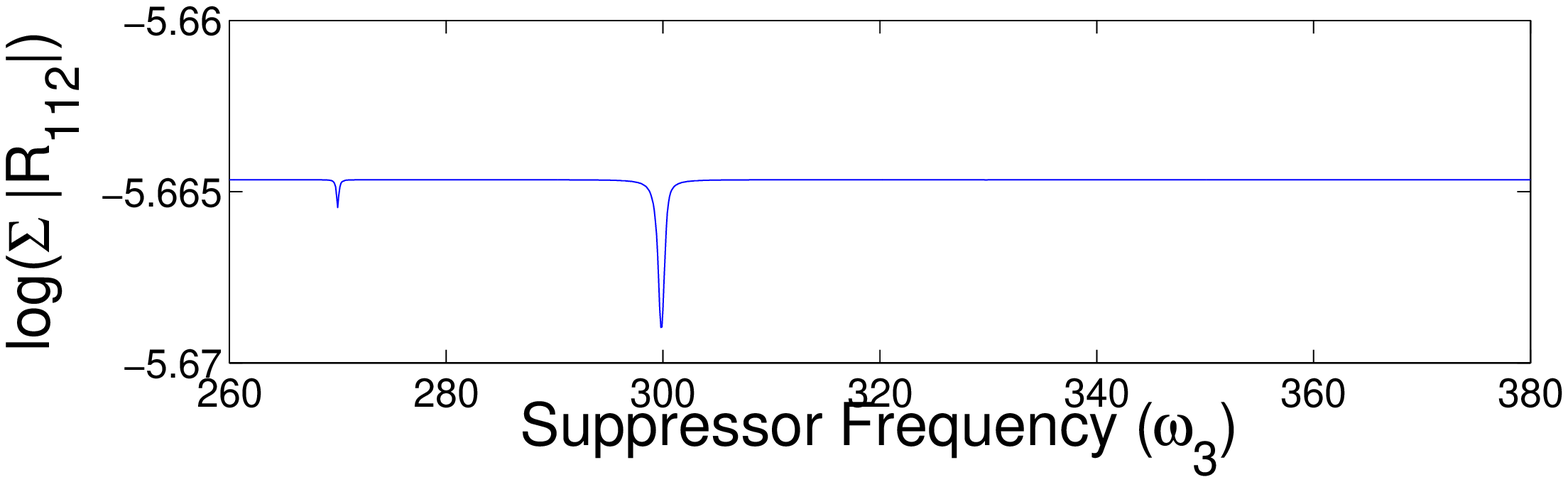}}}

\centerline{(c)}

\centerline{\resizebox{3 in}{!}{\includegraphics{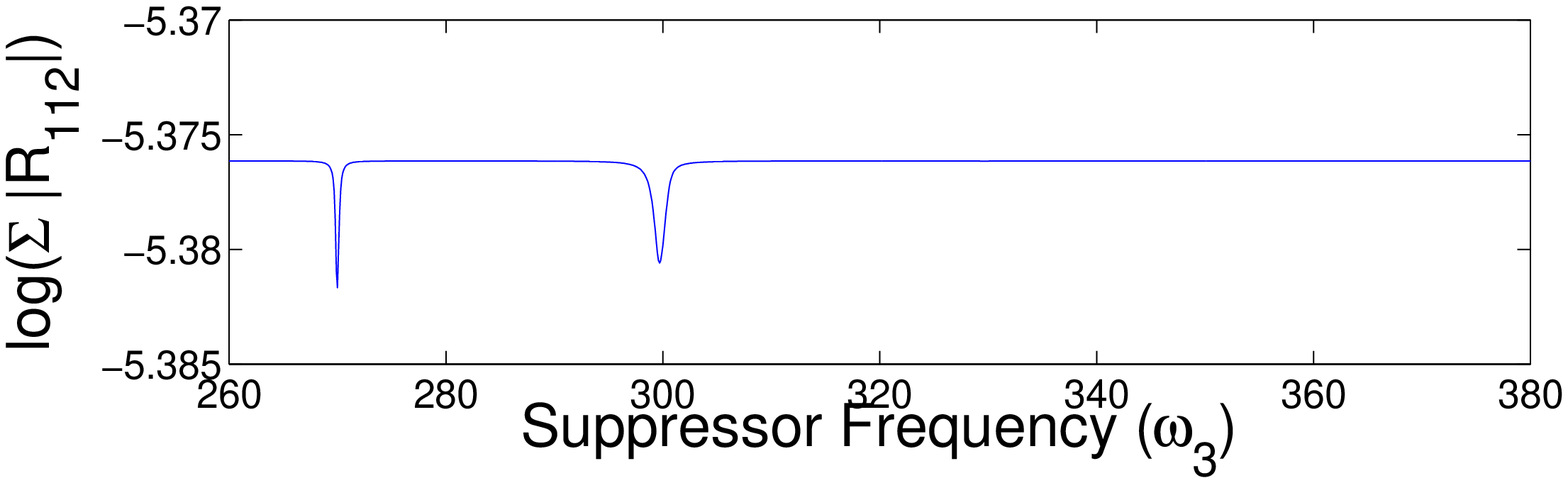}}}

\centerline{(d)}
 
\caption{Figures a.-b. show the magnitude of the response at the $2 \omega_1 - \omega_2$ frequency,  $R_{112}$, for two different cells tuned at different
frequencies as the frequency of the suppressor tone, $\omega_3$  was varied.   For a.-c. parameter values were
set at $a=-.1$, $c=100$, $d=100$, $F_1=.01$, $F_2=.01$, $F_3=.001$.
(a) Cell 1, Natural Frequency, $b=2 \omega_1-\omega_2=270$.
(b) Cell 2, Natural Frequency, $b=\omega_1=300$.
(c) The total $2\omega_1-\omega_2$ component of the response for four cells tuned at 270, 300, 330, and 360.
(d) The total $2\omega_1-\omega_2$ component of the response for the four cells tuned at 270, 300, 330, and 360, with $c=d=500$ and
other parameters the same as a.-c.}
\label{fig:DPcurves1}
\end{figure}

Under the assumption that the cells tuned near the primary frequencies,
$\omega_1$
and $\omega_2$ and the distortion product frequencies,
$2\omega_1-\omega_2$ and $2\omega_2-\omega_1$ are likely 
produce the greatest response at $2\omega_1-\omega_2$, we concentrate on the response of those 
four cells.  Figure \ref{fig:DPcurves1} a.-b. shows the relationship between the
magnitude of the component of the response at the $2 \omega_1-\omega_2$ frequency, $R_{112}$, and
the frequency of the suppressor tone, $\omega_3$, for 
the two cells generated the greatest response, the cells
tuned at $\omega_1=300$ and $2 \omega_1-\omega_2=270$.  
  In figure \ref{fig:DPcurves1} a.-b.,  as observed in the two frequency suppression case, 
maximum suppression occurs when the suppressor frequency is 
tuned near the natural frequency of the cell.  
In this example, the component of the response of the $\omega_1$ 
cell at the distortion product frequency is much louder than the distortion
product component of the response for the other three cells.
So, a plot of the total response of the four cells 
shows that maximum suppression occurs when the suppressor
tone is tuned near $\omega_1$ (figure \ref{fig:DPcurves1} c.).  For larger values
of the forcing frequency, or larger values of the nonlinear
coefficients $c$ and $d$, substantial suppression 
may also occur at the $2\omega_1-\omega_2$ frequency (figure \ref{fig:DPcurves1} d.).
This result is consistent with suppression curve experiments 
in nonmammals which indicate that maximum 
suppression of the response at the distortion product 
frequency, $2 \omega_1 - \omega_2 $, occurs when the suppressor
frequency is near the $\omega_1$ frequency~\cite{PLM91,MK93,MD05,TM98}. Some experiments 
also show a secondary dip near the distortion product frequency, 
as predicted by the model~\cite{TM98}.

\begin{figure}
\centerline{\resizebox{3 in}{!}{\includegraphics{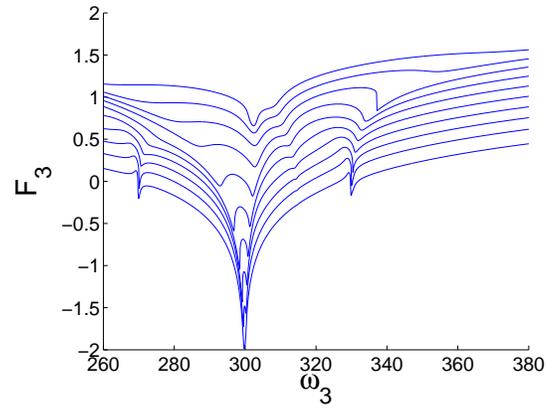}}}

\caption{ Each curve shows the amplitude of the 
suppressor tone, $F_{3}$, needed to suppress the response of a single
Hopf oscillator at the distortion product frequency, $R_{112}$, by a fixed amount. For
the lowest curve in the diagram, the response, $\log(R_{112})$ 
is reduced by $0.5$ from its unsuppressed value. Each consecutive curve shows
the forcing needed to reduce the response by an  
additional 0.5. Parameters were set at $a=-.1$, $b=300$,  $c=100$, $d=100$, $F_1=.01$, and
$F_2=.01$.}
\label{fig:suppression}
\end{figure}

Data recorded in suppression experiments is slightly
different than that shown in figure \ref{fig:DPcurves1} c.-d., where the
forcing amplitude was held constant for each curve.
Typically in suppression experiments, the magnitude of forcing needed to
reduce the component of the response at $2\omega_1 - \omega_2$ by a specified amount is recorded as
the suppressor frequency is changed.  A single Hopf oscillator tuned
at $\omega_1=300$ yields results similar to suppression experiments, again with maximum suppression
occurring near $\omega_1$ (figure \ref{fig:suppression})~\cite{PLM91,MK93,MD05,TM98}. 

While the Hopf oscillator model qualitatively predicts the response properties
for three--frequency suppression experiments in nonmammals,  it does not reproduce
mammalian suppression results.  Recall, in mammals, it is observed that 
maximum suppression of the $2\omega_1-\omega_2$ frequency occurs when 
the suppressor tone is tuned near the $\omega_2$ frequency not the
$\omega_1$ frequency as in nonmammals.  Over many trials, the Hopf oscillator
model never predicted maximum suppression near the $\omega_2$ frequency.
The probable reason for the discrepancy lies in differences in physiology 
between mammals and nonmammals.  In nonmammals, the hair cells are embedded
in a membrane that lacks tuning properties, while in mammals, the hair cells 
are embedded in the basilar membrane~\cite{Kandel}.  The basilar membrane performs
much of the frequency filtering in the mammalian inner ear.  When sound of
a given frequency strikes the inner ear, a traveling wave is set into motion
along the basilar membrane.  This traveling wave reaches its maximum amplitude
at different places along the membrane depending upon the frequency of the
stimulus.  For a high frequency stimulus, the wave reaches its maximum
amplitude closer to the base of the cochlea than it would for lower frequency stimulus.
After the wave passes through its preferred frequency, vibrations at that
frequency are damped.

If the mammalian cochlea is forced at two frequencies, $\omega_1$ and $\omega_2$ with
$\omega_1< \omega_2$, the hair cells tuned near the higher frequency, $\omega_2$ will feel both
frequency components of the stimuli. Because higher frequency stimuli will have 
dissipated by the time the traveling wave reaches the hair cell tuned at $\omega_1$,  that cell will feel
mainly the $\omega_1$ component of the stimulus.  While in nonmammals, the 
cell tuned near $\omega_1$ is responsible for generating the largest portion of
the distortion product otoacoustic emission, in mammals the cell tuned near
the $\omega_1$ frequency does not receive the full stimulus at both frequency components
and cannot produce as great a response at the distortion product frequency.
Therefore, it would not be surprising if most of the $2 \omega_1- \omega_2$   distortion product
frequency was generated at the $\omega_2$ cell and not the $\omega_1$ cell in mammals, causing
maximum suppression to occur near $\omega_2$.

\section*{Conclusions}
\label{sec:conclusions}

A model consisting of a set of noninteracting oscillators tuned near a Hopf bifurcation
was successful in qualitatively predicting the results of three-frequency forcing 
experiments observed in mammals and nonmammals.  In the case of nonmammals, only
two Hopf oscillators tuned near $\omega_1$ and $2\omega_1-\omega_2$ were necessary
to predict the results of three-tone suppression experiments.
In mammals, a Hopf oscillator tuned near the $\omega_2$ frequency
correctly predicted experimental results.  Which
cell contributes the most is dictated by important differences
in mammalian and nonmammalian physiology.  Notably, it was not necessary
to assume coupling between cells of different frequencies in
order to qualitatively reproduce experimental data. Though more complicated biophysically--based 
models would be needed to produce a more quantitative agreement
with the experiments, it is interesting that such a simple model can
explain the main experimental features. These results lend support to
the idea that an array of oscillators tuned near a Hopf bifurcation could be responsible for otoacoustic 
emissions and active sound processing in the ear.  Because both the somatic
motility of the outer hair cell and the motion of the hair bundle are
thought to be modeled by systems poised near a Hopf bifurcation, either could play
the role of the Hopf oscillator.

K.A.M. is grateful for fellowship support through NSF-RTG grant 0354259.

\bibliography{dpoae5bp}

\clearpage

\clearpage

\end{document}